\documentclass[conference]{IEEEtran}
\IEEEoverridecommandlockouts
\usepackage{cite}
\usepackage{amsmath,amssymb,amsfonts}
\usepackage{algorithmic}
\usepackage{graphicx}
\usepackage{textcomp}
\usepackage{multirow}
\usepackage{xcolor}

\newcommand{\eg}  {\emph{e.g., \/}}

\def\BibTeX{{\rm B\kern-.05em{\sc i\kern-.025em b}\kern-.08em
    T\kern-.1667em\lower.7ex\hbox{E}\kern-.125emX}}
\begin{document}

\title{Cyberattack Action-Intent-Framework for Mapping Intrusion Observables\\
\thanks{This research is supported by NSF Awards \#1526383 and \#1742789}
}

\author{\IEEEauthorblockN{Stephen Moskal}
\IEEEauthorblockA{\textit{Department of Computer Engineering} \\
\textit{Rochester Institute of Technology}\\
Rochester, NY, USA \\
sfm5015@rit.edu}
\and
\IEEEauthorblockN{Shanchieh Jay Yang}
\IEEEauthorblockA{\textit{Department of Computer Engineering} \\
\textit{Rochester Institute of Technology}\\
Rochester, NY, USA \\
jay.yang@rit.edu}
}

\maketitle

\begin{abstract}
The techniques and tactics used by cyber adversaries are becoming more sophisticated, ironically, as defense getting stronger and the cost of a breach continuing to rise.  
Understanding the thought processes and behaviors of adversaries is extremely challenging as high profile or even amateur attackers have no incentive to share the trades associated with their illegal activities.
One opportunity to observe the actions the adversaries perform is through the use of Intrusion Detection Systems (IDS) which generate alerts in the event that suspicious behavior was detected.
The alerts raised by these systems typically describe the suspicious actions via the form of attack `signature', which do not necessarily reveal the true intent of the attacker performing the action.
Meanwhile, several high level frameworks exist to describe the sequence or chain of action types an adversary might perform. These frameworks, however, do not connect the action types to observables of standard intrusion detection systems, nor describing the plausible intents of the adversarial actions.
To address these gaps, this work proposes the Action-Intent Framework (AIF) to complement existing Cyber Attack Kill Chains and Attack Taxonomies.
The AIF defines a set of Action-Intent States (AIS) at two levels of description: the Macro-AIS describes `what' the attacker is trying to achieve and the Micro-AIS describes ``how" the intended goal is achieved.
A full description of both the Macro is provided along with a set of guiding principals of how the AIS is derived and added to the framework.
\end{abstract}


\section{Introduction}
A necessary component to next generation cyber attack detection or prediction algorithms is understanding the tactics and the techniques that adversaries use to learn about the network, compromise assets, and eventually achieve their end-objective (stealing information, disrupting services, etc.).
Behaviors of cyber attackers are extremely diverse due to differences in skill level and/or situational behaviors given the target's network configuration.
It is unlikely that two attackers will share the exact same attack processes. 
Efficient extraction and comprehension of attack actions and the corresponding behaviors can enable rapid response or even application to similar network assets for proactive cyber defense.
A common framework that describes a process of attacker actions is the Cyber Attack Kill Chain which describes the attack process as a chain of action types.
Cyber defense may detect intrusion in the earlier stage of the kill chain and `cut the chain' to stop the attacker from performing the later, more critical kill chain stages.
Since the introduction of the Cyber Attack Kill Chain by Lockheed Martin in 2011 \cite{KC_Lockheed}, organizations have adapted the concept of a Kill Chain for specific attack types such as Advanced Persistent Threats (APT), insider threats, etc \cite{KC_Dell}.
Typical kill chains summarize the entire attack sequence typically in less than 10 stages creating a concise and easy to understand attacker description for both security professionals and the average person.
However, when it comes to generally classifying attack actions, most kill chain descriptions are not well suited to differentiate the different tactics and techniques that an adversary may use as they are intended to represent milestones in the attack sequence but not the individual actions.  
Moskal \cite{moskal2016knowledge} gave a comparative study of the different kill chains and summarized the various attack stages into reconnaissance, exploitation, and exfiltration categories. 
Table \ref{kc_counts} shows the number of classes for some of the most popular kill chains.

\begin{table}[h]
\caption{Attack stage classifications by various organizations.}
\begin{tabular}{|l|l|l|l|}
\hline
\multicolumn{1}{|c|}{}   & \multicolumn{1}{c|}{Year} & \multicolumn{1}{c|}{Type} & \multicolumn{1}{c|}{\# of Classes} \\ \hline
MTIRE CAPEC \cite{barnum2008common}             & 2007                      & Attack Patterns           & 517                                \\ \hline
Lockheed Martin \cite{KC_Lockheed}          & 2011                      & Kill Chain                & 7                                  \\ \hline
STIX \cite{KC_Stix}                    & 2012                      & Attack Descriptor        & 9                                  \\ \hline
MITRE Kill Chain         & 2013                      & Kill Chain                & 7                                  \\ \hline
MTIRE Att\&ck \cite{mitreattk}           & 2013                      & Attack Techniques         & 295                                \\ \hline
Varonis \cite{KC_Varonis}                  & 2018                      & Kill Chain                & 8                                  \\ \hline
MITRE Unified Kill Chain \cite{strom2018mitre} & 2018                      & Kill Chain                & 20                                 \\ \hline
\end{tabular}
\label{kc_counts}
\end{table}

MITRE ATT\&CK \cite{strom2018mitre} uses 12 tactics to classify 295 attack technique classes and is an industry-leading and comprehensive attack-type description.
However, it only considers the `right' side of the kill chain: exploitation and exfiltration and the 295 techniques can be too detailed to explicitly map to intrusion observables (\eg IDS alerts, access logs, attack traces, etc.).
This limitation inhibits the modeling and discovery of adversary behaviors and processes. 
Some attack techniques in ATT\&CK are specific to the target network, where the adversary chose the technique because it was easily accessible in that network.
In a different network, the same attacker may exercise the same thought process, come to a similar attack action with the same intended outcome, yet it leads to a different ATT\&CK technique.

Observing the attacker's actions is extremely challenging as it is unlikely that the attacker will cooperate to resolve their intentions and thought processes.
Instead, we turn to the most common and abundant sources of attacker's actions, Intrusion Detection System (IDS) logs, where network traffic passing though a sensor is analyzed against known signatures of specific behaviors raising alerts to the administrators when suspicious actions are performed on the network. 
IDS's are not yet able to resolve the observed actions to the ATT\&CK techniques due to the complexity of IDS rules. 
Instead, we propose the Action-Intent Framework (AIF) containing a set of Action-Intent States (AIS) that is designed to resolve the essence of `what' the attacker was trying to achieve and `how' they achieved it in reference to open-source IDS (\eg Suricata and Snort) signatures.  
To capture `what' the attacker has achieved, we define the concept of Macro-AIS as a high level description of the outcome of the action such as privilege escalation or data destruction.
The Macro-AIS's are similar to the higher-level tactics described in ATT\&CK but also includes reconnaissance stages and zero-day attack stages.
For each Macro-AIS, we define a set of Micro-AIS's that describe `how' the adversary achieves the Macro-AIS.
The Micro-AIS's are similar to the techniques defined by ATT\&CK but with a focus on action-types that are observable by an IDS and not specific to any service, operating system, or network configuration.

\section{Action-Intent State Selection Methodology}
\begin{figure}[h!]
    \centering
    \includegraphics[width=7cm]{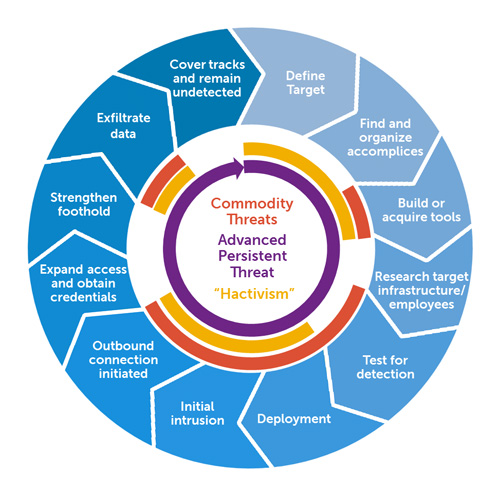}
    \caption{Dell Secureworks Advanced Persistent Threat kill chain stage description}
    \label{fig:dell}
\end{figure}

Kill Chains as a part of their description imply an order of the stages to progress towards a goal and typically are specific to a type of attack/attacker.  
Figure \ref{fig:dell} shows the relevant stages for an attacker depends on the type of attacker or skill level and the cyclical visualization implies that the adversary needs to go through multiple iterations to achieve their end goal.
Our end objective is to use our defined attack stages to discover how the attack stages are used and relate to previous attack stages and upcoming attack stages and because of this we do not define a specific order of stages.
It is this order is what we want to discover and analyze the reasoning the attacker chose these stages and why the particular order is chosen. 

Taking inspiration from MITRE Att\&ck where the ``techniques" are classified under a high-level description called a ``tactic", we employ a similar two-tier structure with a more restrictive and specialized criteria to better capture the intent of the action performed.  
Our Macro AIS definitions contains commonalities to the Att\&ck Tactics but the Macro AIS focuses on resolving action-intents from the defense's perspective where tactics such as ``persistence" or ``defense evasion" describes intentions that may not be possible to definitively determine with just IDS logs.  
The Micro AIS definition also has similarities to the Att\&ck Techniques however a key difference is that the Micro AIS does not include techniques that are specific to any one system such as Kerberoasting or very specialized techniques such as exfiltration though ``Audio Capture".
With these restrictions, we propose a framework to define a set of AIS so that a sequence of AIS derived from IDS observables describes the type of actions specific attacker takes that is general to a network configuration.  

To develop the AIF we define a set of guiding principals that dictate when an action or action type is added to the AIF with a corresponding macro or micro AIS.  
We intend that the AIF will evolve over time (as with MITRE Att\&ck) and changes to the criteria or additional states may be defined in the future.   

\subsection{Macro Action-Intent States}
Table \ref{table:macro} contains the currently defined Macro AIS and their descriptions.
Each of these Macro AIS describes a high level description of ``what" the attacker has performed but does not describe a specific method of to achieve the state.  
Our guiding principals for defining an Macro AIS is as follows:

\begin{enumerate}
    \item The stage describes the impact or end goal of the action,
    \item the stage does not describe a specific means of achieving a goal, and
    \begin{itemize}
        \item For example, there are multiple methods to achieve ``privilege escalation" which has a specific impact.
    \end{itemize}
    \item if an action type has technical or behavioral properties in-which other macro stages do not accurately describe the action's outcome due to different means of observability
    \begin{itemize}
        \item  ``Active reconnaissance" actions (e.g. network scanning) typically can be observed using an IDS where ``passive reconnaissance" like social engineering cannot be observed by traditional technical methods and are used in different situations.
    \end{itemize}
\end{enumerate}

States such as passive and active reconnaissance describe actions where the primary goal is to gather information about the target whether its through publically accessible means (passive) or technical approaches using scanning tools for example (active).
Privilege Escalation, Targeted Exploits, Ensure Access, and Zero-Day describe some sort of exploitation of the target to allow the attacker to gain access or ensure access to the target.  
Although Zero-days are by nature difficult to detect, these types of actions are included as the usage of zero days is a critical differentiator of attacker behaviors and certain types of sensors like anomaly detectors do have the capability to observe zero-days.
States such as Disrupt, Distort, Destroy, Disclosure, and Delivery describes a specific end impact that an attacker may perform as either a sub-goal or end-goal of the overall attack. 
For example attackers may choose to disrupt specific machines as they are traversing the network to draw attention away from the action where crucial information such as customer data is disclosed to the attacker.  
Under each Macro AIS will be a set of Micro AIS which will describe ``how" the attacker chose to achieve the behavior described in as Micro AIS in the following section. 

\begin{table*}[tb!]
\centering
\caption{The Macro Action-Intent States currently defined in the framework.}
\begin{tabular}{|l|l|}
\hline
\multicolumn{1}{|c|}{Macro AIS} & \multicolumn{1}{c|}{Description}                                                                                                                                                          \\ \hline
Passive Recon                                & \begin{tabular}[c]{@{}l@{}}An attempt to gain information about targeted computers \\ and networks without actively engaging with the systems\end{tabular}                                \\ \hline
Active Recon                                 & \begin{tabular}[c]{@{}l@{}}An intruder engages with the targeted system to gather information \\ about vulnerabilities\end{tabular}                                                       \\ \hline
Privilege Escalation.                               & \begin{tabular}[c]{@{}l@{}}Act of exploiting a bug, design flaw or configuration oversight in an \\ operating system or software application to gain elevated access\end{tabular}         \\ \hline
Targeted Exploits                            & \begin{tabular}[c]{@{}l@{}}Exploits targeting a specific service, application, organizational entity, \\ or person\end{tabular}                                                          \\ \hline
Ensure Access                                & \begin{tabular}[c]{@{}l@{}}Actions which expand preexisting access or circumvent active defense\\ strategies\end{tabular}                                                                 \\ \hline
Zero Day                                     & \begin{tabular}[c]{@{}l@{}}Actions performed employing undocumented vulnerabilities or strategies \\ with unknown consequences where no patch exists at the time of the attack\end{tabular} \\ \hline
Disrupt                                     & Disruption in services, usually from a Denial of Service.                                                                                                                                 \\ \hline
Destroy                                      & \begin{tabular}[c]{@{}l@{}}Destruction of information, usually when an attack has caused a deletion of \\ files or removal of access.\end{tabular}                                        \\ \hline
Distort                                      & Distortion in information, usually when an attack has caused a modification of a file.                                                                                                    \\ \hline
Disclosure                                   & \begin{tabular}[c]{@{}l@{}}Disclosure of information, usually providing an attacker with a view of information\\ they would normally not have access to\end{tabular}                      \\ \hline
Delivery                                     & \begin{tabular}[c]{@{}l@{}}Actions where the intent is to place/install/deliver data that could be in the form of\\ malware, backdoor, application, etc.\end{tabular}                     \\ \hline
\end{tabular}
\label{table:macro}
\end{table*}

\subsection{Micro Action-Intent States}
The Micro AIS is similar to the MITRE Att\&ck techniques but with the key difference is that we do not include techniques that are specific to any one service, operating system, or network.  
In the case of privilege escalation, MITRE Att\&ck defines techniques like ``Sudo" and ``Bypass User Account Control" which are legitimate techniques to gain root access to Linux and Windows machines respectively however the choice of choosing these one of these techniques is situational depending on the network.
The true intention of these techniques was to gain administrative access to the machine regardless of the network, thus our methodology is to combine these two techniques based on the intent creating the Micro Action-Intent state of ``Root Privilege Escalation" and also ``User Privilege Escalation".
These are two states with specific impacts to the target, used in different situations, and describes the intentions of the adversary without requiring information about the target network.
Our currently defined Micro AIS is defined in Table \ref{table:micro} and our guiding principals for selecting new states for the Micro AIS is defined below:

\begin{enumerate}
    \item The state describes a specific and unique means of achieving an macro attack stage type,
    \begin{itemize}
        \item Network sniffing credential access and brute force credential access can achieve privilege escalation however are used in different situations and are observed differently
    \end{itemize}
    \item the state is service and platform agnostic,
    \item the state has a well defined impact on a type of target or yields different response from the target, and 
    \begin{itemize}
        \item For example, ``End-point DoS" is used to target services where ``Network DoS" targets a whole network which has significantly different impact 
        \item Attackers may use different types of reconnaissance actions to reveal different characteristics of the network, services, or vulnerabilities of a network
    \end{itemize}
    \item if the state has observable characteristics that differentiates itself from other stages within the same macro attack stage.
    \begin{itemize}
        \item The micro states for ``End-Point DoS" and ``Service Stop" both are disrupting the function of a single machine however end-point DoS implies disruption by exhausting system resources for example and service stop involves directly terminating the process.
    \end{itemize}
\end{enumerate}

\section{Uses for the Action-Intent Framework}
The AIF was designed with a few use cases in mind and may be adapted to a specific use case if needed.  
Given its focus on observable characteristics of an adversary, we believe the AIF can be used for adversarial sequence extraction from IDS alerts as shown by Moskal et al. \cite{moskal2018isi}.
Sequences of common alert type where developed and compared to other attackers to assess commonalities between attacker processes, however the alert signature type used was not reflective of the intentions of the adversary.
We believe that this work can be used to enhance the contextual meaning of these sequences so that the intent of the adversarial actions are easily interpreted.
This concept requires a method to classify alert signatures to the AIF which is not an easy task given the many different types of observables and the vast number of alert rules and architectures available; a future work will be presented on this topic at a later date.
Once sequences of AIS are determined, we also believe that the extracted processes of attacker actions can be simulated where a cyber attack simulator can use the AIS as guidance to select an action that is both representative of the attacker action and what is possible on the target network \cite{moskal2018cyber}.

\begin{table*}[b]
\caption{Descriptions of the Micro Action-Intent States in the current framework with corresponding Macro AIS.}
\begin{tabular}{|c|l|l|}
\hline
Macro AIS                 & \multicolumn{1}{c|}{Micro AIS}                                      & \multicolumn{1}{c|}{Description}                                                                                                                                                                                                      \\ \hline
\multirow{3}{*}{Passive Recon}     & Target Identification                                                        & Determining the organizational/network target                                                                                                                                                                                         \\ \cline{2-3} 
                                   & Surfing                                                                      & \begin{tabular}[c]{@{}l@{}}Using legitimate methods (websites, public documents, etc) to obtain information \\ about the target\end{tabular}                                                                                           \\ \cline{2-3} 
                                   & Social Engineering                                                           & \begin{tabular}[c]{@{}l@{}}Non-technical strategy cyber attackers use that relies heavily on human interaction \\ and often involves tricking people into breaking standard security practices\end{tabular}                           \\ \hline
\multirow{3}{*}{Active Recon}      & Host Discovery                                                               & Use of technical programs to uncover the location/IP of machines in the target network                                                                                                                                                \\ \cline{2-3} 
                                   & Service Discovery                                                            & Use of technical programs to uncover the services or applications employed on a machine                                                                                                                                               \\ \cline{2-3} 
                                   & Vulnerability Discovery                                                      & Techniques or programs to uncover vulnerabilities on machine with a specific application or OS                                                                                                                                        \\ \cline{2-3} 
                                   & Information Discovery                                                        & \begin{tabular}[c]{@{}l@{}}Actions which reveal technical information such as system configurations, file system contents, \\ or information about the target/entity\end{tabular}                                                     \\ \hline
\multirow{5}{*}{Privledge Esc.}    & User Privledge Esc.                                                          & Action which results in the adversary gaining user privileges                                                                                                                                                                          \\ \cline{2-3} 
                                   & Root Privledge Esc.                                                          & Action which results in the adversary gaining root/admin privileges                                                                                                                                                                   \\ \cline{2-3} 
                                   & \begin{tabular}[c]{@{}l@{}}Network Sniffing\\ Credential Access\end{tabular} & \begin{tabular}[c]{@{}l@{}}Using the network interface on a system to monitor or capture information sent over a wired \\ or wireless connection\end{tabular}                                                                         \\ \cline{2-3} 
                                   & \begin{tabular}[c]{@{}l@{}}Brute Force \\ Credential Access\end{tabular}     & \begin{tabular}[c]{@{}l@{}}Brute force techniques to attempt access to accounts when passwords are unknown or when\\ password hashes are obtained\end{tabular}                                                                        \\ \cline{2-3} 
                                   & Account Manipulation                                                         & \begin{tabular}[c]{@{}l@{}}Modifying permissions, modifying credentials, adding or changing permission groups,\\ modifying account settings, or modifying how authentication is performed\end{tabular}                                \\ \hline
\multirow{8}{0pt}{Targeted Exploits} & Trusted Orginization Exploitation                                            & \begin{tabular}[c]{@{}l@{}}Access through trusted third party relationship exploits an existing connection that may\\ not be protected or receives less scrutiny than standard mechanisms of gaining access to a network\end{tabular} \\ \cline{2-3} 
                                   & Exploit Public Facing Application                                            & \begin{tabular}[c]{@{}l@{}}Use of software, data, or commands to take advantage of a weakness in an Internet-facing computer \\ system or program in order to cause unintended or unanticipated behavior\end{tabular}                 \\ \cline{2-3} 
                                   & Exploit Remote Services                                                      & \begin{tabular}[c]{@{}l@{}}Exploitation of remote services such as VPNs, Citrix, and other access mechanisms allow users to\\ connect to internal enterprise network resources from external locations\end{tabular}                   \\ \cline{2-3} 
                                   & Spearphishing                                                                & \begin{tabular}[c]{@{}l@{}}An email spoofing attack that targets a specific organization or individual, seeking unauthorized \\ access to sensitive information\end{tabular}                                                          \\ \cline{2-3} 
                                   & Service-Specific Exploitation                                                & Use of a exploit/vulnerability specific to a system OS, application, and version                                                                                                                                                      \\ \cline{2-3} 
                                   & Arbitrary Code Execution                                                     & control over an target by establishing a communication channel between adversary and target                                                                                                                                           \\ \hline
\multirow{3}{0pt}{Ensure Access}     & Defense Evasion                                                              & Techniques an adversary may use to evade detection or avoid other defenses                                                                                                                                                            \\ \cline{2-3} 
                                   & Command \& Control                                                           & Control over an target by establishing a communication channel between adversary and target                                                                                                                                           \\ \cline{2-3} 
                                   & Lateral Movement                                                             & \begin{tabular}[c]{@{}l@{}}Techniques that enable an adversary to access and control remote systems on a network and could,\\ but does not necessarily, include execution of tools on remote systems\end{tabular}                     \\ \hline
\multirow{3}{*}{Zero Day}          & Privledge Esc.                                                               & Undocumented action that raises the privilege level of the adversary                                                                                                                                                                  \\ \cline{2-3} 
                                   & Targeted Exploit                                                             & Usage of a unpatched and possibly undocumented targeted exploit                                                                                                                                                                       \\ \cline{2-3} 
                                   & Ensure Access                                                                & Unknown method to evade detection or controlling method                                                                                                                                                                               \\ \hline
\multirow{4}{0pt}{Disrupt}          & End Point DoS                                                                & \begin{tabular}[c]{@{}l@{}}Exhausting the system resources those services are hosted on or exploiting the system to cause \\ a persistent crash condition\end{tabular}                                                                \\ \cline{2-3} 
                                   & Network DoS                                                                  & Exhaust the network bandwidth services rely on                                                                                                                                                                                        \\ \cline{2-3} 
                                   & Service Stop                                                                 & Stop or disable services on a system to render those services unavailable to legitimate users                                                                                                                                         \\ \cline{2-3} 
                                   & Resource Hijacking                                                           & \begin{tabular}[c]{@{}l@{}}Leverage the resources of co-opted systems in order to solve resource intensive problems which\\ may impact system and/or hosted service availability\end{tabular}                                         \\ \hline
\multirow{2}{0pt}{Destroy}           & Data Destruction                                                             & \begin{tabular}[c]{@{}l@{}}Destroy data and files on specific systems or in large numbers on a network to interrupt \\ availability to systems, services, and network resources\end{tabular}                                          \\ \cline{2-3} 
                                   & Content Wipe                                                                 & \begin{tabular}[c]{@{}l@{}}Erase the contents of storage devices on specific systems as well as large numbers of systems \\ in a network to interrupt availability to system and network resources\end{tabular}                       \\ \hline
\multirow{3}{0pt}{Distort}           & Data Encryption                                                              & \begin{tabular}[c]{@{}l@{}}Encrypt data on target systems or on large numbers of systems in a network to interrupt\\ availability to system and network resources\end{tabular}                                                        \\ \cline{2-3} 
                                   & Defacement                                                                   & Modify visual content available internally or externally to an enterprise network.                                                                                                                                                    \\ \cline{2-3} 
                                   & Data Manipulation                                                            & Insert, delete, or manipulate data at rest in order to manipulate external outcomes or hide activity.                                                                                                                                 \\ \hline
Disclosure                         & Data Exfiltration                                                            & \begin{tabular}[c]{@{}l@{}}Techniques and attributes that result or aid in the adversary removing files and information from \\ a target network\end{tabular}                                                                         \\ \hline
Delivery                           & Data Delivery                                                                & Intent to place/install/deliver data that could be in the form of malware, backdoor, application, etc.                                                                                                                                \\ \hline
\end{tabular}
\label{table:micro}
\end{table*}

\bibliographystyle{./bibliography/IEEEtran}
\bibliography{./bibliography/smoskal_main.bib}

\end{document}